\documentclass[aps,preprint,prl,superscriptaddress]{revtex4}

\usepackage{graphicx}
\usepackage{amsmath}
\usepackage{amssymb}
\usepackage{units}
\usepackage{color}

\usepackage{setspace}

\begin{document}

\title{Electronic-structural Dynamics in Graphene}

\author{Isabella Gierz}
\email{Isabella.Gierz@mpsd.mpg.de}
\affiliation{Max Planck Institute for the Structure and Dynamics of Matter, Center for Free Electron Laser Science, Hamburg, Germany}
\author{Andrea Cavalleri}
\affiliation{Max Planck Institute for the Structure and Dynamics of Matter, Center for Free Electron Laser Science, Hamburg, Germany}
\affiliation{Department of Physics, Clarendon Laboratory, University of Oxford, Oxford, United Kingdom}

\date{\today}

\begin{abstract}
We review our recent time- and angle-resolved photoemission spectroscopy experiments, which measure the transient electronic structure of optically driven graphene. For pump photon energies in the near infrared ($\hbar\omega_{\text{pump}}=950$meV) we have discovered the formation of a population-inverted state near the Dirac point, which may be of interest for the design of THz lasing devices and optical amplifiers. At lower pump photon energies ($\hbar\omega_{\text{pump}}<400$meV), for which interband absorption is not possible in doped samples, we find evidence for free carrier absorption. In addition, when mid-infrared pulses are made resonant with an infrared-active in-plane phonon of bilayer graphene ($\hbar\omega_{\text{pump}}=200$meV), a transient enhancement of the electron-phonon coupling constant is observed, providing interesting perspective for experiments that report light-enhanced superconductivity in doped fullerites in which a similar lattice mode was excited. All the studies reviewed here have important implications for applications of graphene in optoelectronic devices and for the dynamical engineering of electronic properties with light.
\end{abstract}

\maketitle

\section{Introduction}

Dynamical engineering of the electronic structure of solids with light is emerging as an important new area of research, which complements more traditional routes of control using chemistry, pressure, or magnetic fields. Especially interesting are new techniques that can drive materials close to lattice and other resonances at mid-infrared and THz frequencies. Indeed, in these driven solids new properties emerge away from equilibrium. These result naturally from modulation with a strong, periodic field that creates new effective Hamiltonians with new eigenstates \cite{PolkovnikovRevModPhys2011}.
 
Resonant excitation of the crystal lattice has, for example, been used to induce superconductivity far above the equilibrium critical temperature in cuprates \cite{FaustiScience2011, KaiserPRB2014, HuNatMater2014, HuntPRB2015} and K$_3$C$_{60}$ \cite{MitranoNature2016}, to induce insulator-to-metal phase transitions \cite{RiniNature2007, CavigliaPRL2012, HuPRB2016}, and to melt magnetic or orbital order \cite{FörstPRB2011, CavigliaPRB2013, TobeyPRL2008}. Recently, the formation of photon-dressed Floquet-Bloch states upon mid-infrared pumping has been shown to result in a topological phase transition in Bi$_2$Se$_3$ \cite{WangScience2013, MahmoodNatPhys2016}. Although these techniques have been primarily applied to strongly correlated materials, the underlying mechanism is more general and should be studied in simpler conditions.

Here we set out to study the dynamical electronic properties of monolayer and bilayer graphene in three different excitation regimes. (1) For pump photon energies bigger than twice the chemical potential, electron-hole pairs are generated. Photo-excited electrons are found to rapidly accumulate above the Dirac point at the bottom of the conduction band where they form a short-lived population-inverted state (Fig. \ref{fig1}) \cite{Gierz_popinv}. By improving the temporal resolution of the experiment to better than 10fs we are able to resolve the primary scattering events that are responsible for the formation of the population inversion (Fig. \ref{fig2}) \cite{Gierz_10fs,Gierz_NTCs}. (2) For lower pump photon energies free carriers in the vicinity of the Fermi level are accelerated by the applied laser field, and then heated via inelastic scattering. For this excitation regime, we observe a thermal electronic distribution at all times within the experimental energy and time resolution (Fig. \ref{fig3}) \cite{Gierz_popinv}. (3) When the pump frequency is tuned to be resonant to the infrared-active in-plane E$_{\text{1u}}$ phonon in bilayer graphene at E$_{\text{ph}}$=200meV, in addition to carrier heating via free carrier absorption, the electronic band structure is modulated by the coherent oscillation of the carbon atoms. In this case we observe anomalous carrier dynamics that we attribute to a transient enhancement of the electron-phonon coupling constant (Fig. \ref{fig4}) \cite{Gierz_phonon,Gierz_dekinking}.

\section{Methods}

\subsection{Samples}

For the present study we used epitaxial monolayer and bilayer graphene samples on hydrogen-terminated silicon carbide crystals. The 6H-SiC(0001) substrates were hydrogen-etched and annealed in Argon atmosphere, resulting in carbon coverages of one and two monolayers, respectively. These carbon layers were then decoupled from the underlying substrate by hydrogen intercalation, giving quasi-freestanding monolayer and bilayer graphene \cite{RiedlPRL2009}. The equilibrium band structure of these samples can be found in \cite{Gierz_phonon}. For all samples the top of the valence band is unoccupied, indicating hole-doping with the chemical potential at about -200meV below the Dirac point. The samples were exposed to air and reinserted into ultra-high vacuum at the Artemis user facility in Harwell, United Kingdom, where the tr-ARPES experiments were performed. The original band structure was recovered by a mild annealing.

\subsection{Setup}

Tr-ARPES measurements have been performed at the Artemis user facility at Harwell, United Kingdom. Femtosecond laser pulses tunable between 800nm and 15$\mu$m wavelength are typically used to excite the sample. At a variable delay from this excitation, the sample is illuminated with a second extreme ultra-violet (XUV) probe pulse that ejects photoelectrons from the sample. These photoelectrons are measured with a hemispherical analyzer, yielding the photocurrent as a function of kinetic energy, E$_{\text{kin}}$, and emission angle, $\theta$. From the kinetic energy, the binding energy, $E_{\text{B}}$, of the electrons in the solid can be obtained, and the measured emission angle can be converted into in-plane momentum, k$_{||}$, so that the image on the two-dimensional detector directly maps the band structure along a particular cut in momentum space.

The setup is based on a 30fs/780nm/1kHz Titanium:Sapphire (Ti:Sa) laser. Pump pulses are generated with a commercial optical parametric amplifier (1 to 2$\mu$m) with subsequent difference frequency generation (4 to 15$\mu$m). About 1mJ of laser energy is used for high-order harmonics generation in an Argon gas jet, generating XUV photons in the energy range between 10 and 40eV. A single-grating time-preserving monochromator is used to select a single harmonic at $\hbar\omega_{\text{probe}}$=30eV, yielding $<$30fs long XUV pulses with a spectral width of about 300meV \cite{Frassetto2011}. The XUV probe is focused onto the sample with a toroidal mirror resulting in a typical spot size of 200$\mu$m.
 
In order to improve the temporal resolution of the setup to below 10fs, the output spectrum of the Ti:Sa laser was broadened in a Neon filled fiber and recompressed using chirped mirrors. The compressed pulses were used both for high order harmonics generation and to pump the sample \cite{Gierz_10fs}.

The photocurrent was measured along the $\Gamma$K-direction in the vicinity of the K-point as a function of momentum, energy, and pump-probe delay. Electron distribution functions have been obtained by integrating the data over momentum \cite{Gierz_popinv, Ulstrup_2014, Gierz_phonon}.

\section{Results}

\subsection{Interband excitation}

\begin{figure}
	\center
  \includegraphics[width = 1\columnwidth]{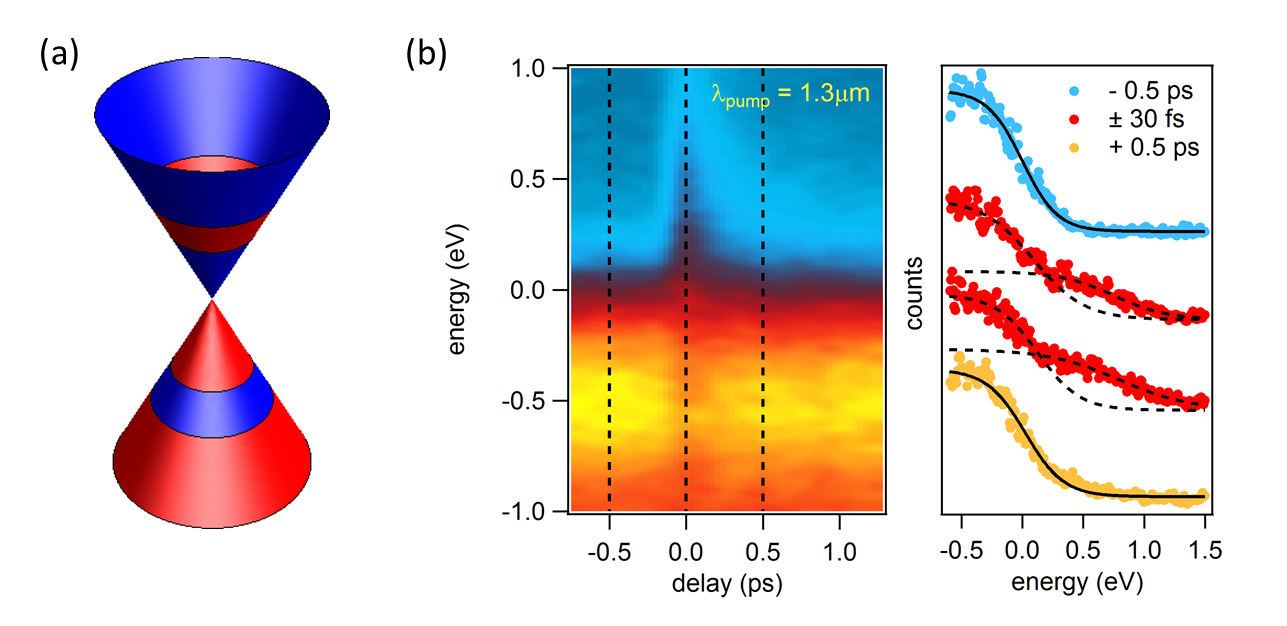}
  \caption{(a) Schematic of direct interband excitation. (b) Momentum-integrated photocurrent in the vicinity of the K-point in monolayer graphene as a function of delay (left) together with distribution functions at selected pump-probe delays (right) for $\hbar\omega_{\text{pump}}$ = 950meV.}
  \label{fig1}
\end{figure}

The interband excitation regime is discussed in detail in Fig. \ref{fig1}. In Fig. \ref{fig1}a we show a schematic of the excitation mechanism. Measured electron distributions for various pump-probe delays before, during, and after excitation are shown in Fig. \ref{fig1}b \cite{Gierz_popinv}. At negative time delays, before the arrival of the pump pulse, the measured electron distributions (blue data points in Fig. \ref{fig1}b) follow a Fermi-Dirac distribution (continuous black lines). Upon arrival of the pump pulse, however, clear deviations from a single Fermi-Dirac function develop (red data points in Fig. \ref{fig1}b). In this case, the data can be fitted with two Fermi functions, one with the chemical potential inside the valence band and another one with the chemical potential inside the conduction band (dashed black lines in Fig. \ref{fig1}b). The two chemical potentials merge within $\sim$130fs and a single Fermi-Dirac distribution is recovered (yellow data points in Fig. \ref{fig1}b). We attribute the transient distribution of electrons described by two separate chemical potentials located in the valence and conduction band, respectively, to a population-inverted state \cite{Gierz_popinv}.

The measurements displayed in Fig. \ref{fig1} show a thermal electron distribution at all times, either described by a single or two separate Fermi-Dirac distributions for valence and conduction band. This indicates that the temporal resolution of these measurements ($\sigma$=35fs, FWHM=80fs, determined from the width of the rising edge of the pump probe signal) is too bad to allow access to the primary thermalization events. 

\subsection{Extreme Timescales}

\begin{figure*}
	\center
  \includegraphics[width = 1\columnwidth]{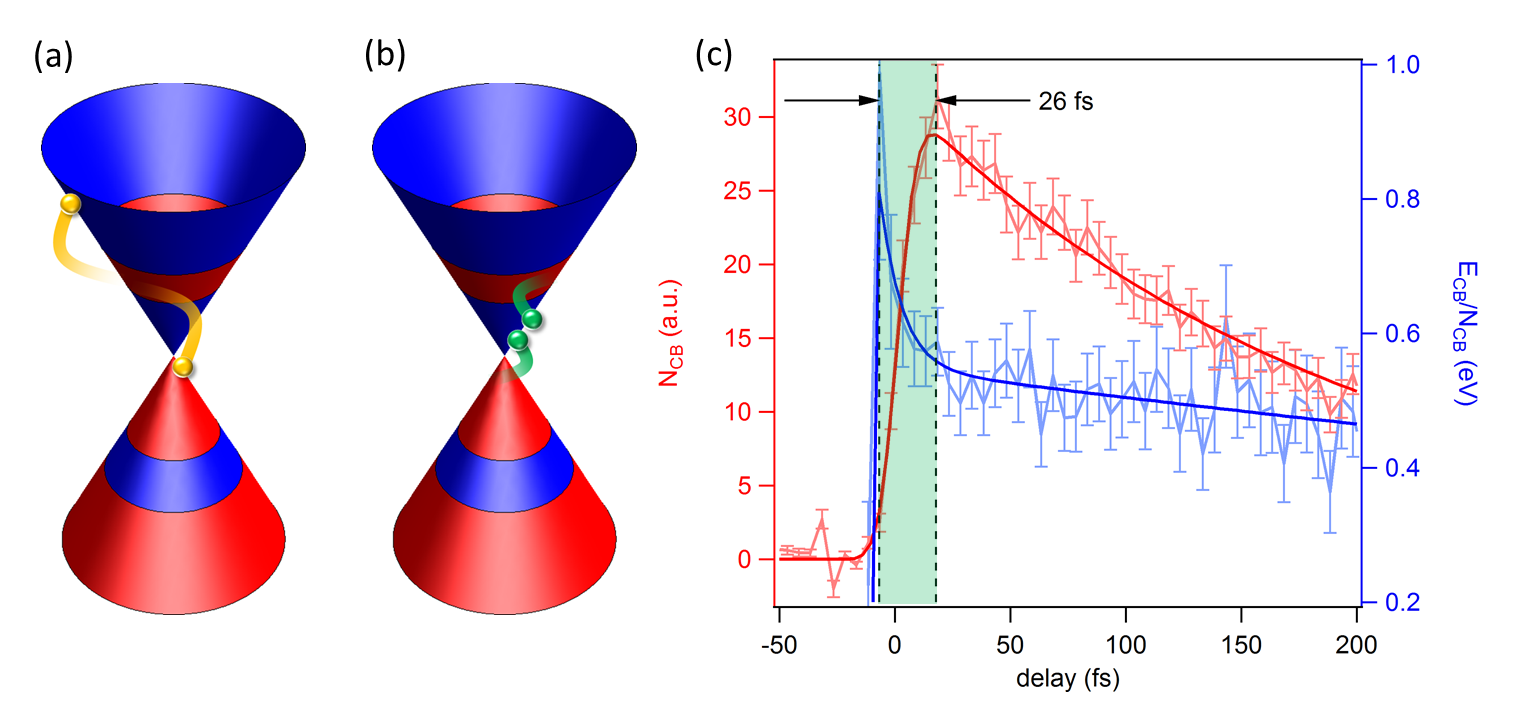}
  \caption{Primary thermalization events tracked down with sub 10 fs pulses. (a) Schematic of Auger heating. This process is expected to be strongly suppressed in graphene due to the lack of empty (occupied) states at the top (bottom) of the valence (conduction) band. (b) Schematic of impact ionization. The presence of occupied (empty) states at the top (bottom) of the valence (conduction) band results in a large phase space for impact ionization. (c) Comparison of the total number of carriers inside the conduction band N$_{\text{CB}}$ and their average kinetic energy E$_{\text{CB}}$/N$_{\text{CB}}$ as a function of delay. For about 25 fs after photoexcitation N$_{\text{CB}}$ keeps increasing while E$_{\text{CB}}$/N$_{\text{CB}}$ already decreases, indicating impact ionization.}
  \label{fig2}
\end{figure*}

The ultrafast $<$30fs dynamics after photo-excitation are predicted to be dominated by different Auger processes as sketched in Fig. \ref{fig2}a and b \cite{WinzerNanoLett2010, WinzerPRB2012}. A photo-excited electron-hole pair can recombine and transfer its energy to a second electron inside the conduction band (Fig. \ref{fig2}a). This process (Auger recombination) is dominant in conventional semiconductors with a parabolic dispersion \cite{Beattie1958, Svantesson1971, Auston1975, Benz1976}. In graphene, due to the conical dispersion and the absence of a band gap, this process is strongly suppressed by the lack of occupied (empty) states at the bottom (top) of the conduction (valence) band. Instead, carrier relaxation is expected to be dominated by impact ionization (Fig. \ref{fig2}b), where the excess energy of an electron high up in the conduction band is used to generate a second electron-hole pair \cite{WinzerNanoLett2010, WinzerPRB2012, PlötzingNanoLett2014}. 

These two processes can be easily identified by measuring the number of carriers inside the conduction band, N$_{\text{CB}}$, as well as their average kinetic energy, E$_{\text{CB}}$/N$_{\text{CB}}$. In the case of Auger recombination, N$_{\text{CB}}$ decreases while E$_{\text{CB}}$/N$_{\text{CB}}$ increases and vice versa for impact ionization. These numbers are directly accessible in a tr-ARPES experiment, as the photocurrent is proportional to the number of electrons at any given energy \cite{Gierz_10fs}.

In order to resolve these primary scattering events, we repeated the experiment with a temporal resolution of $\sigma$=9fs (FWHM=21fs) using a pump photon energy of $\hbar\omega_{\text{pump}}$=1.6eV. Figure \ref{fig2}c shows both N$_{\text{CB}}$ as well as E$_{\text{CB}}$/N$_{\text{CB}}$ as a function of pump-probe delay. We find that, during the first $\sim$25fs after photo-excitation, N$_{\text{CB}}$ keeps increasing while E$_{\text{CB}}$/N$_{\text{CB}}$ already decreases, clearly indicating impact ionization \cite{Gierz_10fs}.

\subsection{Free Carrier Absorption}

\begin{figure*}
	\center
  \includegraphics[width = 1\columnwidth]{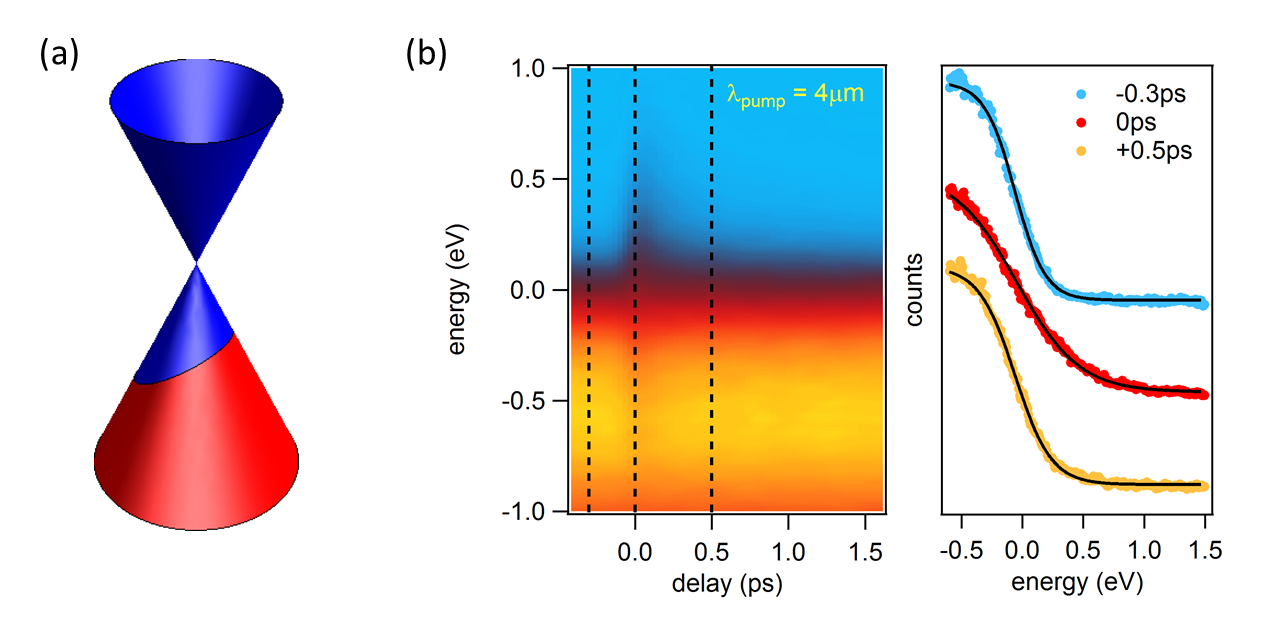}
  \caption{(a) Schematic of free carrier absorption. (b) Momentum-integrated photocurrent in the vicinity of the K-point in monolayer graphene as a function of delay (left) together with distribution functions at selected pump-probe delays (right) for $\hbar\omega_{\text{pump}}$ = 300meV. }
  \label{fig3}
\end{figure*}

For experiments in which the pump photon energy was below the threshold for interband excitation, metallic carriers in our hole-doped samples absorb energy from the pump by phonon-assisted intraband absorption (Fig. \ref{fig3}a). This excitation mechanisms dominates the pump-probe signal in our samples if $\hbar\omega_{\text{pump}}<$400meV. In Fig. \ref{fig3}b we show electron distributions for different delays after excitation at $\hbar\omega_{\text{pump}}$=300meV. We find that a thermal electronic distribution is maintained at all times within the experimental resolution. The only effect of the pump pulse is a broadening of the distribution, indicating an elevated electronic temperature \cite{Gierz_popinv}. 

\subsection{Phonon Pumping}

\begin{figure*}
	\center
  \includegraphics[width = 1\columnwidth]{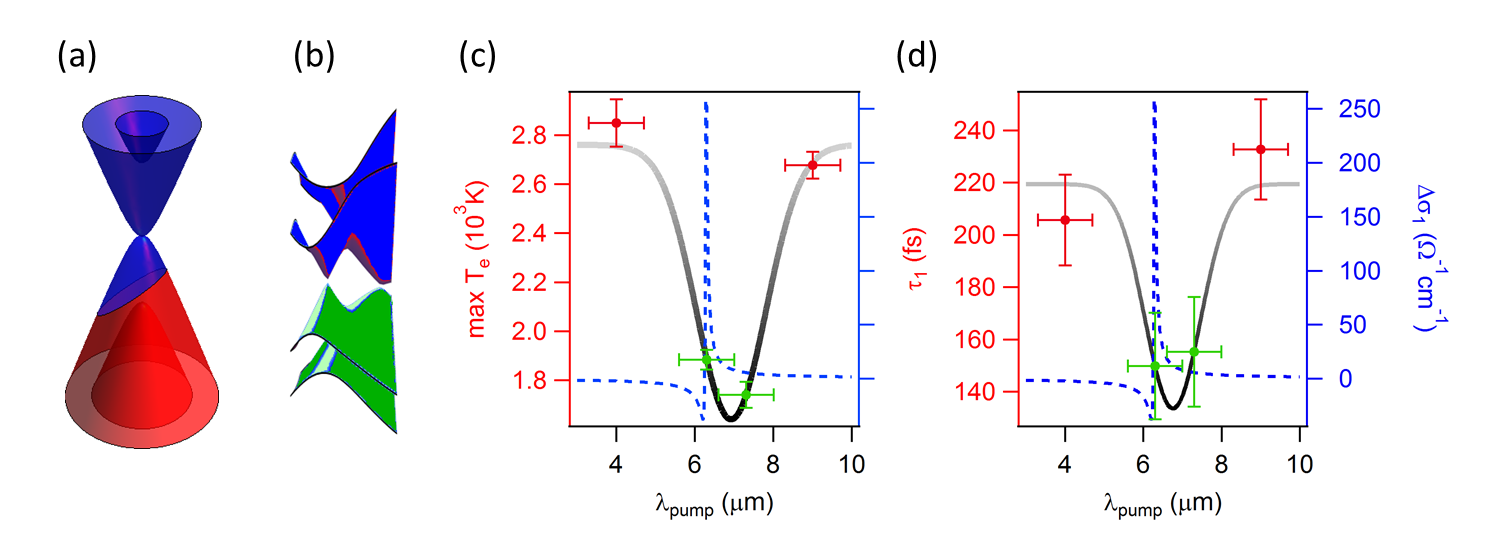}
  \caption{Low-energy excitations of bilayer graphene. (a) Schematic of free carrier absorption in bilayer graphene. This process dominates carrier heating if $\hbar\omega_{\text{pump}}<2|\mu_e|$. (b) Resonant excitation of the in-plane bond-stretching phonon at $\hbar\omega_{\text{pump}}$ = 200meV periodically splits the Dirac cone along the $\Gamma$K-direction. (c) Peak electronic temperature as a function of pump wavelength. (d) Initial fast cooling time attributed to the emission of strongly coupled optical phonons as a function of pump-probe delay. Green (red) data points are for on- (off-) resonance excitation. Measurements in (c) and (d) were performed for a constant fluence of 0.26mJ/cm$^2$. Dashed blue lines in (c) and (d) show the optical conductivity associated with the driven phonon from \cite{KuzmenkoPRL2009}.}
  \label{fig4}
\end{figure*}

The excitation schemes discussed above only redistribute the electrons inside the Dirac cone and leave both the single particle dispersion as well as the coupling strength of various many-body interactions unaffected. Therefore, in order to manipulate the electronic pro\-per\-ties with light, one needs to apply different driving schemes. In this section we choose the pump frequency to be resonant with the infrared-active in-plane bond stretching E$_{\text{1u}}$ phonon in bilayer graphene at E$_{\text{ph}}$=200meV. (Monolayer graphene does not possess any infrared-active modes and cannot be used for the present purpose.) In this case, in addition to carrier heating via free carrier absorption (Fig. \ref{fig4}a), the crystal lattice and thus the band structure is periodically modulated (Fig. \ref{fig4}b). Both the electronic peak temperature (Fig. \ref{fig4}c) and the fast cooling time (Fig. \ref{fig4}d), typically interpreted in terms of optical phonon emission, are found to decrease when the pump pulse is tuned to the phonon resonance. We attribute this to a transient increase of the electron-phonon coupling constant \cite{Gierz_phonon, Gierz_dekinking}.

\section{Discussion}

Most of the pump-probe data on graphene is interpreted in terms of a three-temperature model (see e. g. \cite{LuiPRL2010, JohannsenPRL2013}). On time scales short compared to the typical experimental resolution the photo-excited electrons thermalize into a Fermi-Dirac distribution at an elevated electronic temperature, T$_{\text{e}}$ \cite{BreusingPRL2009, BreusingPRB2011}. In a second step, on a timescale on the order of 200fs, the electrons cool down by the emission of a subset of strongly coupled optical phonons, mainly the E$_{\text{2g}}$ mode at q=$\Gamma$ and the A$_1'$ mode at q=K \cite{KampfrathPRL2005, YanPRB2009, KangPRB2010, ChatzakisPRB2011, CalandraPRB2007, ParkNanoLett2008}. Once the electrons and the optical phonons have reached the same temperature, further cooling of the electron-optical-phonon system occurs via emission of acoustic phonons within several picoseconds \cite{KampfrathPRL2005, LuiPRL2010, YanPRB2009, KangPRB2010, ChatzakisPRB2011, SongPRL2012}. Similar models have been applied to describe the cooling dynamics of hot carriers within many different sample systems such as metals \cite{AllenPRL1987, ErnstorferPRX2016}, semimetals \cite{IshidaPRB2016}, semiconductors \cite{DaiPRB2015}, topological insulators \cite{LaiAPL2014}, and high-temperature superconductors \cite{PerfettiPRL2007}.

In this discussion, we want to focus on the peculiarities of photo-excited graphene that go beyond the well-known conclusions obtained from a three-temperature model, i.e., (1) the observation of a transient population inversion, (2) pre-thermal scattering events dominated by impact ionization, and (3) anomalous dynamics due to resonant excitation of the crystal lattice.

(1) and (2) are related to the conical band structure of graphene and the vanishing density of states at the Dirac point. While the occurrence of a population-inverted state after strong optical excitation is commonly observed in conventional semiconductors with a sizable band gap, it is not a priori clear whether such a state might exist in graphene. Ryzhii et al. \cite{RyzhiiJApplPhys2007} predicted that photo-excited electron-hole pairs in graphene relax by optical phonon emission, resulting in an accumulation of electrons (holes) at the bottom (top) of the conduction (valence) band and a negative optical conductivity in the THz regime. The effect has been demonstrated by Li et al. \cite{LiPRL2012} and recently confirmed by more detailed experimental \cite{Gierz_popinv} and theoretical investigations \cite{WinzerPRB2013, JagoPRB2015}. Aside from the interaction with optical phonons, several theoretical \cite{WinzerNanoLett2010, WinzerPRB2012} and experimental papers \cite{PlötzingNanoLett2014} have stressed the importance of different Auger scattering channels in photo-excited graphene.

In our tr-ARPES experiments performed in different excitation regimes and with different temporal resolutions we are able to observe the complete dynamics during the buildup and decay of the population inversion \cite{Gierz_NTCs}. From Fig. \ref{fig2} we know that impact ionization rapidly accumulates the photo-excited electrons at the bottom of the conduction band, assisting the buildup of the population-inversion via optical phonon emission \cite{Gierz_10fs}. While two separate Fermi-Dirac distributions can be clearly distinguished for about 130fs in Fig. \ref{fig1}b \cite{Gierz_popinv}, complete thermalization of the carriers into a single Fermi-Dirac distribution takes about 250fs \cite{Gierz_NTCs}. The decay of the population inversion within hundreds of femtoseconds is dominated by Auger recombination with a minor contribution of optical phonon emission \cite{WinzerPRB2013}.

The observed anomalous carrier dynamics that occur when the E$_{\text{1u}}$ phonon is resonantly excited can be understood in terms of a transiently enhanced electron-phonon coupling constant \cite{Gierz_phonon, Gierz_dekinking}. In Fig. \ref{fig4}c and d we observe lower peak electronic temperatures as well as faster cooling rates at resonance with the E$_{\text{1u}}$ mode \cite{Gierz_phonon}. A more detailed analysis of the energy-dependence of the scattering rates shows that resonant excitation of the phonon results in systematically higher scattering rates at all energies compared to off-resonance excitation \cite{Gierz_dekinking}. In \cite{Gierz_dekinking} we simulated the coupled electron-phonon dynamics using Boltzmann equations and showed that neither the difference in the measured electron distribution nor an enhanced occupancy of the driven phonon mode can reproduce our data, leaving an enhanced electron-phonon coupling constant as the only plausible explanation for our findings. 

This can also be understood from intuitive arguments without the need to invoke theoretical models. An increase in electronic temperature increases the phase space for optical phonon emission and is expected to result in faster cooling of the electronic system \cite{KemperPRB2014}. We, however, observe a faster cooling rate for smaller electronic temperatures (Fig. \ref{fig4}c,d and \cite{Gierz_phonon}). Further, an increased occupation of the driven phonon will enhance the probability for both stimulated phonon emission and phonon absorption. Phonon emission (carrier cooling) only dominates over phonon absorption (carrier heating) in the presence of a population-inverted electronic distribution. We observe a thermal electronic distribution at all times before, during, and after excitation in the mid-infrared \cite{Gierz_popinv, Gierz_phonon}. Finally, in the framework of a two-temperature model, an increased cooling rate at smaller temperature difference between electrons and phonons can only come from an increased coupling strength between the two systems.

\section{Conclusion}

We have reviewed the electronic-structural dynamics of optically excited graphene at different pump wavelengths throughout the near- and mid-infrared. These dynamics were tracked with time- and angle-resolved photoemission spectroscopy. We find that free-carrier absorption efficiently heats the electrons to high temperatures but maintains a thermal electronic distribution at all times. After interband excitation, carriers rapidly accumulate at the bottom of the conduction band, resulting in a short-lived population-inverted state. Using sub 10fs pulses we were able to show that the predominat scattering channel during the first ~25fs after interband excitation is impact ionization. Finally, resonant excitation of the E$_{\text{1u}}$ mode in bilayer graphene leads to faster relaxation at lower peak electronic temperatures that we attribute to a transient enhancement of the electron-phonon coupling constant. 

While the observed short-lived population inversion may in principle be used to amplify TeraHertz pulses, the presence of impact ionization most likely cannot be exploited for efficient light harvesting as the absence of a band gap in graphene makes electron-hole separation difficult.

The observed transient enhancement of the electron-phonon coupling strength presents a significant step towards dynamical band structure engineering with important consequences for the occurrence and the understanding of light-induced superconductivity in graphene and other carbon compounds, such as K$_3$C$_{60}$ \cite{MitranoNature2016}.

\section{Acknowledgments} 

We thank F. Calegari, S. Aeschlimann, M. Ch{\'a}vez-Cervantes, M. Mitrano, C. Cacho, E. Springate, J. Petersen, C. R. Ast, and U. Starke for their contributions to a successful outcome of the experiments and B. Sachs, T. O. Wehling, S. Lichtenstein, M. Eckstein, and M. A. Sentef for theoretical support. The research presented here received funding from Laserlab Europe, STFC, and the German Science Foundation (SPP 1459 and SFB 925).

\clearpage

\begin{thebibliography}{12}

\bibitem{PolkovnikovRevModPhys2011} A. Polkovnikov, K. Sengupta, A. Silva, and M. Vengalattore, Rev. Mod. Phys 83, 863 (2011) 
 \bibitem{FaustiScience2011} D. Fausti, R. I. Tobey, N. Dean, S. Kaiser, A. Dienst, M. C. Hoffmann, S. Pyon, T. Takayama, H. Takagi, A. Cavalleri, Science 331, 189 (2007)
\bibitem{KaiserPRB2014} S. Kaiser, C. R. Hunt, D. Nicoletti, W. Hu, I. Gierz, H. Y. Liu, M. Le Tacon, T. Loew, D. Haug, B. Keimer, and A. Cavalleri, Phys. Rev. B 89, 184516 (2014)
\bibitem{HuNatMater2014} W. Hu, S. Kaiser, D. Nicoletti, C. R. Hunt, I. Gierz, M. C. Hoffmann, M. Le Tacon, T. Loew, B. Keimer and A. Cavalleri, Nat. Mater. 13, 705 (2014)
\bibitem{HuntPRB2015} C. R. Hunt, D. Nicoletti, S. Kaiser, T. Takayama, H. Takagi, and A. Cavalleri, Phys. Rev. B 91, 020505(R) (2015)
\bibitem{MitranoNature2016} M. Mitrano, A. Cantaluppi, D. Nicoletti, S. Kaiser, A. Perucchi, S. Lupi, P. Di Pietro, D. Pontiroli, M. Ricc{\`o}, S. R. Clark, D. Jaksch, and A. Cavalleri, Nature 530, 461 (2016)
\bibitem{RiniNature2007} M. Rini, R. Tobey, N. Dean, J. Itatani, Y. Tomioka, Y. Tokura, R. W. Schoenlein, and A. Cavalleri, Nature 449, 72 (2007)
\bibitem{CavigliaPRL2012} A. D. Caviglia, R. Scherwitzl, P. Popovich, W. Hu, H. Bromberger, R. Singla, M. Mitrano, M. C. Hoffmann, S. Kaiser, P. Zubko, S. Gariglio, J.-M. Triscone, M. F{\"o}rst, and A. Cavalleri, Phys. Rev. Lett. 108, 136801 (2012)
\bibitem{HuPRB2016} W. Hu, S. Catalano, M. Gibert, J.-M. Triscone, and A. Cavalleri, Phys. Rev. B 93, 161107(R) (2016)
\bibitem{FörstPRB2011} M. F{\"o}rst, R. I. Tobey, S. Wall, H. Bromberger, V. Khanna, A. L. Cavalieri, Y.-D. Chuang, W. S. Lee, R. Moore, W. F. Schlotter, J. J. Turner, O. Krupin, M. Trigo, H. Zheng, J. F. Mitchell, S. S. Dhesi, J. P. Hill, and A. Cavalleri, Phys. Rev. B 84, 241104(R) (2011)
\bibitem{CavigliaPRB2013} A. D. Caviglia, M. F{\"o}rst, R. Scherwitzl, V. Khanna, H. Bromberger, R. Mankowsky, R. Singla, Y.-D. Chuang, W. S. Lee, O. Krupin, W. F. Schlotter, J. J. Turner, G. L. Dakovski, M. P. Minitti, J. Robinson, V. Scagnoli, S. B.Wilkins, S. A. Cavill, M. Gibert, S. Gariglio, P. Zubko, J.-M. Triscone, J. P. Hill, S. S. Dhesi, and A. Cavalleri, Phys. Rev. B 88, 220401(R) (2013)
\bibitem{TobeyPRL2008} R. I. Tobey, D. Prabhakaran, A. T. Boothroyd, and A. Cavalleri, Phys. Rev. Lett. 101, 197404 (2008)
\bibitem{WangScience2013} Y. H. Wang, H. Steinberg, P. Jarillo-Herrero, and N. Gedik, Science 342, 453 (2013)
\bibitem{MahmoodNatPhys2016} F. Mahmood, C.-K. Chan, Z. Alpichshev, D. Gardner, Y. Lee, P. A. Lee, and N. Gedik, Nat. Phys. 12, 306 (2016)
\bibitem{Gierz_popinv} I. Gierz, J. C. Petersen, M. Mitrano, C. Cacho, I. C. E. Turcu, E. Springate, A. St{\"o}hr, A. K{\"o}hler, U. Starke, and A. Cavalleri, Nat. Mater. 12, 1119 (2013)
\bibitem{Gierz_10fs} I. Gierz, F. Calegari, S. Aeschlimann, M. Ch{\'a}vez Cervantes, C. Cacho, R.?T. Chapman, E. Springate, S. Link, U. Starke, C.?R. Ast, and A. Cavalleri, Phys. Rev. Lett. 115, 086803 (2015)
\bibitem{Gierz_NTCs} I. Gierz, arXiv:1607.03287 (2016) 
\bibitem{Gierz_phonon} I. Gierz, M. Mitrano, H. Bromberger, C. Cacho, R. Chapman, E. Springate, S. Link, U. Starke, B. Sachs, M. Eckstein, T. O. Wehling, M. I. Katsnelson, A. Lichtenstein, and A. Cavalleri, Phys. Rev. Lett. 114, 125503 (2015)
\bibitem{Gierz_dekinking} I. Gierz, M. Ch{\'a}vez-Cervantes, M. Mitrano, R. Tomar, H. Bromberger, H. Liu, S. Kaiser, M. A. Sentef, A. St{\"o}hr, S. Link, U. Starke, C. Cacho, R. Chapman, E. Springate, F. Frasetto, L. Poletto, and A. Cavalleri, arXiv:1607.02314 (2016)
\bibitem{RiedlPRL2009} C. Riedl, C. Coletti, T. Iwasaki, A. A. Zakharov, and U. Starke, Phys. Rev. Lett. 103, 246804 (2009)
\bibitem{Frassetto2011} F. Frassetto, C. Cacho, C. A. Froud, I. C. E. Turcu, P. Villoresi, W. A. Bryan, E. Springate, and L. Poletto, Opt. Express 19, 19169 (2011)
\bibitem{Ulstrup_2014} S. Ulstrup, J. C. Johannsen, M. Grioni, and P. Hofmann, Rev. Sci. Instr. 85, 013907 (2014)
\bibitem{WinzerNanoLett2010} T. Winzer, A. Knorr, and E. Mali{\'c}, Nano Lett. 10, 4839 (2010)
\bibitem{WinzerPRB2012} T. Winzer, and E. Mali{\'c}, Phys. Rev.  B 85, 241404(R) (2012)
\bibitem{Beattie1958} A. R. Beattie, and P. T. Landsberg, Proc. R. Soc. A 249, 16 (1958)
\bibitem{Svantesson1971} K. G. Svantesson, N. G. Nilsson, and L. Huldt, Solid State Commun. 9, 213 (1971)
\bibitem{Auston1975} D. H. Auston, C. V. Shank, and P. LeFur, Phys. Rev. Lett. 35, 1022 (1975)
\bibitem{Benz1976} G. Benz, and R. Conradt, Phys. Rev. B 16, 843 (1976)
\bibitem{PlötzingNanoLett2014} T. Pl{\"o}tzing, T. Winzer, E. Mali{\'c}, D. Neumaier, A. Knorr, and H. Kurz, Nano Lett. 14, 5371 (2014) 
\bibitem{LuiPRL2010} C. H. Lui, K. F. Mak, J. Shan, and T. F. Heinz, Phys. Rev. Lett. 105, 127404 (2010) 
\bibitem{JohannsenPRL2013} J. C. Johannsen, S. Ulstrup, F. Cilento, A. Crepaldi, M. Zacchigna, C. Cacho, I. C. E. Turcu, E. Springate, F. Fromm, C. Raidel, T. Seyller, F. Parmigiani, M. Grioni, and P. Hofmann, Phys. Rev. Lett. 111, 027403 (2013) 
\bibitem{BreusingPRL2009} M. Breusing, C. Ropers, and T. Elsaesser, Phys. Rev. Lett. 102, 086809 (2009) 
\bibitem{BreusingPRB2011} M. Breusing, S. Kuehn, T. Winzer, E. Mali{\'c}, F. Milde, N. Severin, J. P. Rabe, C. Ropers, A. Knorr, and T. Elsaesser, Phys. Rev. B 83, 153410 (2011) 
\bibitem{KampfrathPRL2005} T. Kampfrath, L. Perfetti, F. Schapper, C. Frischkorn, and M. Wolf, Phys. Rev. Lett. 95, 187403 (2005) 
\bibitem{YanPRB2009} H. Yan, D. Song, K. F. Mak, I. Chatzakis, J. Maultzsch, and T. F. Heinz, Phys. Rev. B 80, 121403(R) (2009) 
\bibitem{KangPRB2010} K. Kang, D. Abdula, D. G. Cahill, and M. Shim, Phys. Rev. B 81, 165405 (2010) 
\bibitem{ChatzakisPRB2011} I. Chatzakis, H. Yan, D. Song, S. Berciaud, and T. F. Heinz, Phys. Rev. B 83, 205411 (2011) 
\bibitem{CalandraPRB2007} M. Calandra, and F. Mauri, Phys. Rev. B 76, 205411 (2007) 
\bibitem{ParkNanoLett2008} C.-H- Park, F. Giustino, M. L. Cohen, and S. G. Louie, Nano Lett. 8, 4229 (2008) 
\bibitem{SongPRL2012} J. C. W. Song, M. Y. Reizer, and L. S. Levitov, Phys. Rev. Lett. 109, 106602 (2012) 
\bibitem{AllenPRL1987} R. B. Allen, Phys. Rev. Lett. 59, 1460 (1987) 
\bibitem{ErnstorferPRX2016} L. Waldecker, R. Bertoni, R. Ernstorfer, and J. Vorberger, Phys. Rev. X 6, 021003 (2016) 
\bibitem{IshidaPRB2016} Y. Ishida, H. Masuda, H. Sakai, S. Ishiwata, and S. Shin, Phys. Rev. B 93, 100302(R) (2016) 
\bibitem{DaiPRB2015} Y. M. Dai, J. Bowlan, H. Li, H. Miao, S. F. Wu, W. D. Kong, P. Richard, Y. G. Shi, S. A. Trugman, J.-X. Zhu, H. Ding, A. J. Taylor, D. A. Yarotski, and R. P. Prasankumar, Phys. Rev. B 92, 161104(R) (2015) 
\bibitem{LaiAPL2014} Y.-P. Lai, H.-J. Chen, K.-H. Wu, and J.-M. Liu, Appl. Phys. Lett. 105, 232110 (2014) 
\bibitem{PerfettiPRL2007} L. Perfetti, P. A. Loukakos, M. Lisowski, U. Bovensiepen, H. Eisaki, and M. Wolf, Phys. Rev. Lett. 99, 197001 (2007) 
\bibitem{RyzhiiJApplPhys2007} V. Ryzhii, M. Ryzhii, and T. Otsuji, J. Appl. Phys. 101, 083114 (2007) 
\bibitem{LiPRL2012} T. Li, L. Luo, M. Hupalo, J. Zhang, M. C. Tringides, J. Schmalian, and J. Wang, Phys. Rev. Lett. 108, 167401 (2012) 
\bibitem{WinzerPRB2013} T. Winzer, E. Mali{\'c}, and A. Knorr, Phys. Rev. B 87, 165413 (2013) 
\bibitem{JagoPRB2015} R. Jago, T. Winzer, A. Knorr, and E. Mali{\'c}, Phys. Rev. B 92, 085407 (2015) 
\bibitem{KemperPRB2014} A. F. Kemper, M. A. Sentef, B. Moritz, J. K. Freericks, and T. P. Devereaux, Phys. Rev. B 90, 075126 (2014) 
\bibitem{KuzmenkoPRL2009} A. B. Kuzmenko, L. Benfatto, E. Cappelluti, I. Crassee, D. van der Marel, P. Blake, K. S. Novoselov, and A. K. Geim, Phys. Rev. Lett. 103, 116804 (2009)

\end{thebibliography}
\end{document}